\documentclass{article}
\usepackage{spconf,amsmath,graphicx}
\usepackage{bm}
\usepackage{amsmath}

\usepackage{changepage}

\title{A generalized framework for domain adaptation of PLDA in Speaker Recognition}
\name{Qiongqiong Wang, Koji Okabe, Kong Aik Lee, Takafumi Koshinaka}
\address{Biometrics Research Laboratories, NEC, Japan}

\begin{document}
%
\maketitle
\begin{abstract}
This paper proposes a generalized framework for domain adaptation of Probabilistic Linear Discriminant Analysis (PLDA) in speaker recognition.
It not only includes several existing supervised and unsupervised domain adaptation methods 
but also makes possible more flexible usage of available data in different domains. 
In particular, we introduce here the two new techniques described below.
(1) Correlation-alignment-based interpolation and 
(2) covariance regularization.
The proposed correlation-alignment-based-interpolation method decreases $\rm min$$\rm C_{primary}$ up to 30.5\% 
as compared with that from an out-of-domain PLDA model before adaptation, 
and $\rm min$$\rm C_{primary}$ is also 5.5\% lower than with a conventional linear interpolation method with optimal interpolation weights. 
Further, the proposed regularization technique ensures robustness in interpolations w.r.t. varying interpolation weights, 
which in practice is essential. 
\end{abstract}


\section{Introduction}
\label{sec:intro}

Recent progress in speaker recognition has achieved successful application of deep neural networks to derive deep speaker embeddings from speech utterances \cite{snyder17,variani14, snyder18, okabe18}. Speaker embeddings are fixed-length continuous-value vectors that provide succinct characterizations of speakers’ voices rendered in speech utterances. 
Similar to classical i-vectors \cite{dehak11},
deep speaker embeddings live in a simpler Euclidean space in which distance can be measured far more easily
than with much more complex input patterns. 
Techniques such as within-class covariance normalization (WCCN) \cite{hatch06}, linear discriminant analysis (LDA) \cite{bishop06},
and probabilistic LDA (PLDA) \cite{prince07, ioffe06, kenny10} can also be applied.

State-of-the-art speaker recognition systems that are composed of an x-vector (or i-vector) speaker embedding front-end followed by a PLDA backend have shown promising performance \cite{yamamoto19}. 
The effectiveness of these components relies on the availability of a large collection of labeled training data, 
typically over hundred hours of speech recordings consisting of multi-session recordings from several thousand speakers.
It would be prohibitively expensive, however, to collect such a large amount of in-domain (InD) data for a new domain of interest for every application. 
Most available resource-rich data that already exist will not match new domains of interest, i.e., most will be out-of-domain (OOD) data. 
The challenge of domain mismatch arises when a speaker recognition system is used in a different domain (e.g., with different languages, demographics etc.)
from that of the training data. 
Performance may degrade considerably.

Domain adaptation techniques for adapting resource-rich OOD systems so as to produce good results in new domains, have recently been studied with the aim of alleviating this problem.
They are either supervised adaptation \cite{aronowitz14, garcia14a, misra14,glembek14}, 
for which a small amount of InD data and their speaker labels are used,
or unsupervised adaptation \cite{shum14,garcia14b,villalba14,garcia14c}, for which InD data is used without speaker labels. 
Supervised domain adaptation is more powerful than unsupervised.

Supervised domain adaptation methods can be further categorized into the three approaches described below:
1) Data pooling. 
It has been proposed, for example, to add InD data to a large amount of OOD data to train PLDA \cite{misra14}.
2) Feature vector compensation. 
Data shifting for OOD data has been proposed that uses statistical information about data in both domains \cite{aronowitz14}.
3) PLDA parameter adaptation. 
A linear interpolation method has been proposed for combining parameters of PLDAs trained separately with OOD and InD data
so as to take advantage of both PLDAs \cite{garcia14a};
in \cite{wang16}, a maximum likelihood linear transformation has been proposed for transforming OOD PLDA parameters so as to be closer to InD.
Among unsupervised methods, there are CORAL \cite{alam18} 
as 2) feature vector compensation 
as well as CORAl+ \cite{lee19b} and clustering methods \cite{shum14} as 3) PLDA parameter adaptation. 

Among the three approaches, PLDA parameter adaptation, such as linear interpolation, has advantages over the others.
First, it directly optimizes the model in an efficient way and does not require computationally expensive retraining with large-scale OOD data or transformation of individual feature vectors.
Secondly, it can easily adjust the mixing rate for OOD and InD data by changing interpolation weights.
Simple linear modeling, however, which implicitly assumes small differences between OOD and InD data, is not always feasible in real-world situations.
Further, if interpolation weights are not appropriately determined, performance may seriously deteriorate.

In this paper, we take advantage of previous work \cite{lee19b} and propose 
1) a correlation-alignment-based interpolation and 
2) a covariance regularization for both unsupervised and supervised methods,
on the basis of linear interpolation \cite{garcia14a} for robust domain adaptation. 
Finally, we combine existing and proposed methods into a generalized framework and demonstrate its use in certain special cases. 
Domain adaptation is successful in all cases. 

The remainder of this paper is organized as follows: 
Section 2 reviews the CORAL+ unsupervised method and linear interpolation supervised methods; 
Section 3 introduces the proposed correlation-alignment-based interpolation and a regularization technique, as well as 
a generalized framework for domain adaptation and examples of its use; 
Section 4 describes our experimental setup, results, and analyses;
and Section 5 summarizes our work. 


\section{DOMAIN ADAPTATIONS FOR PLDA}
\label{sec:domainAdaptation}

\subsection{Probabilistic Linear Discriminant Analysis}
\label{ssec:htcoral}
Let vector $\phi$ be a speaker embedding (e.g., x-vector, ivector, etc.). 
We assume that vector $\phi$ is generated from a linear Gaussian model \cite{bishop06}, as follows \cite{prince07, ioffe06}:
\vspace*{-1mm}
\begin{equation}
p({\phi} | \bm{\mathrm{h}} , \bm{\mathrm{x}}) = \mathcal{N}(\phi | \mu + \bm{\mathrm{Fh} }+ \bm{\mathrm{Gx}}, \bm{\Sigma} )
\vspace*{-1mm}
\end{equation}
Vector $\mu$ represents the global mean, while $\bm{\mathrm{F}}$ and $\bm{\mathrm{G}}$ are, respectively, the speaker and channel loading matrices, 
and the diagonal matrix $\bm{\Sigma}$ models residual variances. 
The variables $\bm{\mathrm{h}}$ and $\bm{\mathrm{x}}$ are, respectively, the latent speaker and channel variables, which can be considered between- and within-speaker covariance matrices:
\vspace*{-1mm}
\begin{equation}
\bm {\mathrm {\Phi_b }}= \bm {\mathrm {{FF}}}^ \mathrm {T},
\bm {\mathrm {\Phi_w}} = \bm {\mathrm {{GG}}}^ \mathrm {T} +\bm { \Sigma}.
\vspace*{-1mm}
\end{equation}

PLDA adaptation involves the adaptation of its mean vector and covariance matrices. 
Mean shift due to domain mismatch could be dealt with by centralizing the datasets to a common origin \cite{lee17}.
In this paper, we focus on the adaptation of between- and within-speaker covariance in PLDA. 

\subsection{CORAL+ Unsupervised Method}
\label{ssec:coral}
CORAL+ \cite{lee19b} is a correlation alignment based model-level unsupervised domain adaptation. 
It adapts both the between- and within- speaker covariance matrices given only the total covariance matrix estimated directly from InD data.

Pseudo-InD between- and within- speaker covariance matrices ${\bf{\Phi}}_{\mathrm{I,pseudo}} $ are first computed from pseudo-InD data that is recolored from whitened OOD vectors,
using the total covariance matrices of OOD 
and InD vectors $\{ {\bf{C}}_\mathrm{O}, {\bf{C}}_\mathrm{I}\}$ \cite{alam18}. 
It is commonly known that a linear transformation on a normally
distributed vector leads to an equivalent transformation
on the mean vector and covariance matrix of its density function.
Thus, 
\vspace*{-1mm}
\begin{equation}
{\bf{\Phi}}_{\mathrm{I,pseudo}}
={\bf{C}}_{\mathrm{I}} ^{1/2} {\bf{C}}_{\mathrm{O}} ^{-1/2} 
{\bf{\Phi}}_{\mathrm{O}} 
{\bf{C}}_{\mathrm{O}} ^{-1/2} {\bf{C}}_{\mathrm{I}} ^{1/2},
\label{eq:pseudo}
\vspace*{-1mm}
\end{equation}
where 
${\bf{\Phi}}_{\mathrm{O}}$ is the covariance matrix of the OOD PLDA.
In CORAL+, the adapted PLDA covariance matrices ${\bf{\Phi^+}}$ are: 
\vspace*{-1mm}
\begin{equation}
{\bf{\Phi^+}}= \beta {\bf{\Phi}}_{\mathrm{O}} 
+ (1-\beta)
\Gamma_{max} ( 
{\bf{\Phi}}_{\mathrm{I,pseudo}},
{\bf{\Phi}}_{\mathrm{O}}
) 
\label{eq:coralplus}
\vspace*{-1mm}
\end{equation}
Here, $\Gamma_{max} (\bf {Y},\bf {Z})$ is a regularization function that ensures that the variance increases by choosing the larger value between two covariance matrices $\bf \{E, I\}$ in a diagonalized space after a transformation with matrix $\bf B$ as:
%
\vspace*{-1mm}
\begin{equation}
\begin{split}
\Gamma_{max}({\bf {Y}}, {\bf {Z}})=
{\bf{B}} ^ {\mathrm {-T}} 
max(\bf{ E, I}) {\bf{ B}} ^{-1},
\label{eq:max} \\
{\bf{B}}^{\mathrm {-T}} \bf Y \bf{B} = E,
{\bf{B}}^{\mathrm {-T}} \bf Z \bf{B} = I.
\end{split}
\vspace*{-2mm}
\end{equation}
%
Here, $max(.)$ is a element-wise operator.

The effect of CORAL+ has been experimentally validated on the recent NIST 2016 and 2018 Speaker Recognition Evaluation (SRE’16, SRE’18) datasets \cite{lee19b} \cite{lee19a}.

\subsection{Linear Interpolation Supervised Method}
\label{ssec:lip}

Major features in CORAL+ adaptation include (1) CORAL transformation, (2) covariance regularization, and (3) linear interpolation. 
When the first two factors are dropped, the adapted equation in ~\eqref{eq:coralplus} is reduced to linear interpolation. 
It has been shown in \cite{garcia14a} that linear interpolation, though simple, is a promising method. 
It employs a linear combination, with a weight $\alpha$, of PLDA parameters, i.e., between- and within-speaker covariance of independently trained OOD and InD PLDAs: 
\vspace*{-2mm}
\begin{equation}
\vspace*{-1mm}
{\bf{\Phi^+}}= \alpha {\bf{\Phi}}_{\mathrm{I}} 
+ (1-\alpha)
{\bf{\Phi}}_{\mathrm{O}}
,
\label{eq:lip}
\vspace*{-1mm}
\end{equation}
where $ {\bf{\Phi}}_{\mathrm{I}} $ represents the InD covariance matrix.

Linear interpolation \cite{garcia14a} implicitly assumes, however, that simple interpolation is sufficient, and
such an assumption may not hold if the characteristics of OOD and InD are significantly different. 
In addition, performance is strongly affected by interpolation weights. 

\section{Proposed method}
\label{sec:propose}

%
In this section, we utilize the advantage of CORAL+ \cite{lee19b} and 
propose (1) correlation-alignment-based interpolation (CIP) and (2) covariance regularization used in both unsupervised and supervised methods
for robust domain adaptation. 
Finally, we propose a generalized framework.

\subsection{Correlation-Alignment-Based Interpolation}
\label{ssec:cip}
%
As noted in the introduction, linear interpolation~\cite{garcia14a} assumes that 
no two domains to be interpolated should be significantly distant from one another. 
If OOD PLDA could be transformed into something that is closer to a true InD PLDA, 
we would be able to make the resulting PLDA more reliable. 
For this reason, we propose replacing the OOD covariance matrix in linear interpolation with that of a pseudo-InD PLDA obtained by correlation alignment (CORAL):
\vspace*{-1mm}
\begin{equation}
{\bf{\Phi^+}}= \alpha {\bf{\Phi}}_{\mathrm{I}} 
+ (1-\alpha)
{\bf{\Phi}}_{\mathrm{I,pseudo}}
\label{eq:cip}
\vspace*{-1mm}
\end{equation}
%
%
As noted in Section~\ref{ssec:coral}, CORAL aims to align covariance matrices so that they will match the InD feature space, while maintaining the good properties that OOD PLDA learned from a large amount of data. 
We refer to the process represented in \eqref{eq:cip} as correlation-alignment-based interpolation (CIP).

\subsection{Covariance Regularization Technique}
\label{sssec:reg}
The central idea in domain adaptation is to propagate the uncertainty seen in the InD data to the PLDA model. 
Neither the adaptation equation for linear interpolation (LIP) in~\eqref{eq:lip}, 
nor that for correlation-alignment-based interpolation (CIP) in~\eqref{eq:cip},
guarantees that the variance, and therefore the uncertainty, will increase. 
To deal with this, we introduce covariance regularization that will guarantee an increase in variance.
Here,
LIP with regularization (LIP reg) is given by
\vspace*{-1mm}
\begin{equation}
{\bf{\Phi^+}}= \alpha {\bf{\Phi}}_{\mathrm{I}} 
+ (1-\alpha )
\Gamma_{max} ( 
{\bf{\Phi}}_{\mathrm{O}},
{\bf{\Phi}}_{\mathrm{I}}
) ,
\label{eq:lipreg}
\vspace*{-1mm}
\end{equation}
while CIP with regularization (CIP reg) is
\begin{equation}
{\bf{\Phi^+}}= \alpha {\bf{\Phi}}_{\mathrm{I}} 
+ (1-\alpha )
\Gamma_{max} ( 
{\bf{\Phi}}_{\mathrm{I,pseudo}},
{\bf{\Phi}}_{\mathrm{I}}
). 
\label{eq:cipreg}
\end{equation}
Also, note that the $\Gamma_{max}$ operator is the same as that in \eqref{eq:max}.

\subsection{A Generalized Framework for PLDA Adaptation}
\label{ssec:generalform}

In previous sections, we have presented various adaptation equations that work with covariance matrices.
Three main factors are
(1) interpolations of covariance matrices, 
(2) correlation alignment, and 
(3) covariance regularization.
The adaptations \eqref{eq:coralplus}, \eqref{eq:lip}, \eqref{eq:cip}, \eqref{eq:lipreg}, and \eqref{eq:cipreg}
could be summarized with a single formula as follows:
%
\vspace*{-1mm}
\begin{equation}
{\bf{\Phi^+}}= \alpha {\bf{\Phi}}_{\mathrm{0}} 
+ (1-\alpha)
\Gamma_{max} ( 
{\bf{\Phi}}_{\mathrm{1}},
{\bf{\Phi}}_{\mathrm{2}}
) ,
\label{eq:general}
\vspace*{-1mm}
\end{equation}
where
$ {\bf{\Phi}}_{\mathrm{0}} $ is the covariance matrix of a base PLDA from which a new PLDA is adapted;
${\bf{\Phi}}_{\mathrm{1}}$ is the covariance matrix of a developer PLDA that is supposed to have some properties 
that are the same as or similar to the actual InD PLDA;
${\bf{\Phi}}_{\mathrm{2}}$ is the covariance matrix of a reference PLDA for comparison with the developer PLDA covariance matrix. 
The three PLDAs can be the same or different. 
When the same model is chosen for ${\bf{\Phi}}_{\mathrm{1}}$ and ${\bf{\Phi}}_{\mathrm{2}}$, \eqref{eq:general} will be equivalent to that without regularization.

The above-mentioned domain adaptation methods, as well as a KALDI unsupervised domain adaptation method \cite{kaldi}, can be formulated in a single generalized framework with specific covariance matrices as parameters (see Table~\ref{tab:generalform1}).

\begin{table}[h] 
\begin{adjustwidth}{-0.1cm}{}
\bgroup
\def\arraystretch{1.1}%
\setlength\tabcolsep{2.0pt}
\caption{The special cases derived from the general form.}
\begin{tabular}{c|*{4}{l}r}
\hline
& Method & $\Phi_{0}$ & $\Phi_{1}$ & $\Phi_{2}$ & Eq. \\
\hline \hline
1 & CORAL+\cite{lee19b} & ${\bf{\Phi}}_{\mathrm{O}}$ & ${\bf{\Phi}}_{\mathrm{I,pseudo}}$ & ${\bf{\Phi}}_{\mathrm{O}}$ &\eqref{eq:coralplus}\\
2 & KALDI\cite{kaldi} & ${\bf{\Phi}}_{\mathrm{O}}$ &$ {\bf{C}}_\mathrm{O}$ & ${\bf{\Phi}}_{\mathrm{O}}^b + {\bf{\Phi}}_{\mathrm{O}}^w $ & - \\
3 & LIP \cite{garcia14a} & ${\bf{\Phi}}_{\mathrm{I}}$ & ${\bf{\Phi}}_{\mathrm{O}}$ & ${\bf{\Phi}}_{\mathrm{O}}$ & \eqref{eq:lip} \\
4 & LIP reg & ${\bf{\Phi}}_{\mathrm{I}}$ & ${\bf{\Phi}}_{\mathrm{O}}$ & ${\bf{\Phi}}_{\mathrm{I}}$ & \eqref{eq:lipreg} \\
5 & CIP & ${\bf{\Phi}}_{\mathrm{I}}$ & ${\bf{\Phi}}_{\mathrm{I,pseudo}}$ & ${\bf{\Phi}}_{\mathrm{I,pseudo}}$ & \eqref{eq:cip} \\
6 & CIP reg & ${\bf{\Phi}}_{\mathrm{I}}$ & ${\bf{\Phi}}_{\mathrm{I,pseudo}}$ & ${\bf{\Phi}}_{\mathrm{I}}$ & \eqref{eq:cipreg} \\
7 & - & ${\bf{\Phi}}_{\mathrm{I}}$ &${\bf{\Phi}}_{\mathrm{I,pseudo}}$ & ${\bf{\Phi}}_{\mathrm{O}}$ & \eqref{eq:case7} \\
8 & - &${\bf{\Phi}}_{\mathrm{I}}$ & $\Gamma_{max}({\bf{\Phi}}_{\mathrm{I,pseudo}},{\bf{\Phi}}_{\mathrm{O}})$ & ${\bf{\Phi}}_{\mathrm{I}}$ & \eqref{eq:case8}\\
\hline
\end{tabular}
\label{tab:generalform1}
\egroup
\end{adjustwidth}
\vspace*{-4mm}
\end{table}


In addition, there are more cases that can be derived from the generalized framework.
For example, Special Case 7 in Table~\ref{tab:generalform1} is a variation of CIP with regularization (CIP reg) as 
\vspace*{-1mm}
\begin{equation}
{\bf{\Phi^+}}= \alpha {\bf{\Phi}}_{\mathrm{I}} 
+ (1-\alpha )
\Gamma_{max} ( 
{\bf{\Phi}}_{\mathrm{I,pseudo}},
{\bf{\Phi}}_{\mathrm{O}}
). 
\label{eq:case7}
\vspace*{-1mm}
\end{equation}
Rather than using InD PLDA as a reference for covariance regularization, Special Case 7 uses OOD PLDA. 
Special Case 8 is another variation of CIP reg that employs pseudo-InD PLDA regularized using OOD PLDA as the developer covariance: 
\vspace*{-1mm}
\begin{equation}
{\bf{\Phi^+}}= \alpha {\bf{\Phi}}_{\mathrm{I}} 
+ (1-\alpha )
\Gamma_{max} ( 
\Gamma_{max} ( 
{\bf{\Phi}}_{\mathrm{I,pseudo}},
{\bf{\Phi}}_{\mathrm{O}}
),
{\bf{\Phi}}_{\mathrm{I}}
). 
\label{eq:case8}
\vspace*{-1mm}
\end{equation}
Note that Special Cases1 and 2 are unsupervised methods, while Case 3 to 8 are supervised methods.

\section{Experiments}
\label{sec:exp}
Experiments were conducted on the recent SRE’18 dataset. 
Performance was evaluated in terms of equal error rate (EER) and minimum detection cost ($\rm min$$\rm C_{primary}$) \cite{sre18}.

The latest SREs organized by NIST have focused on domain mismatch as a particular technical challenge.
Switchboard, VoxCeleb 1 and 2, and MIXER corpora that consisted of SREs 04 – 06, 08, 10, and 12 were used to train an x-vector extractor. 
They were considered to be OOD data as they are English speech corpora,
while SRE'18 is in Tunisian Arabic.
Data augmentation applied to the OOD data follows that of our work in \cite{lee19a}.

Only the MIXER corpora and its augmentation, which consisted of 262,427 segments from 4,322 speakers in total, was used as OOD data in PLDA training.
SRE'18 has three datasets: 
an evaluation set (13,451 segments), 
a development set (1,741 segments), 
and an unlabeled set (2,332 segments). 
We chose the bigger labeled dataset, the evaluation set, as IND data to train the PLDA, 
and we conducted an evaluation on the development set. 
The enroll and test data in this section are those in the development set. 
The unlabeled set is used for adaptive symmetric score normalization \cite{colibro17} in all the experiments.


The x-vector extractor is a 43-layers TDNN with residual connections and 
a 2-head attentive statistics pooling in the same way as in \cite{lee19a}.
The number of dimensions of the x-vector was 512. 
Mean shift was applied to OOD data using its mean. 
InD data and enroll and test data were centralized using InD data.
As is commonly done in most state-of-the-art systems, LDA was used to reduce dimensionality to 150-dimension. 
In our interpolation domain adaptation experiments, LDA trained with OOD data was applied to both InD and OOD vectors for training and evaluations.
For the single InD PLDA training, LDA trained with InD data was used. 


\subsection{Results and Analysis}
\label{ssec:results}
We first evaluated 
%
%
performance of the PLDA trained using either OOD or InD data, respectively (see Table~\ref{tab:exp1}).
Also shown in the table are the OOD PLDA adapted to InD in an unsupervised manner, 
i.e., using InD data without labels, with CORAL+~\cite{lee19b},
and supervised manner with LIP~\cite{garcia14a}. 
The weights in both adaptations were chosen to be $0.5$. 
\begin{table}[t] 
\centering
\caption{PLDA with CORAL+ and linear interpolation domain adaptations.}
\begin{tabular}[t]{l|cc}
\hline
Systems & EER(\%) & $\rm min$$\rm C_{primary}$ \\
\hline \hline
InD PLDA &4.15 &0.293 \\
OOD PLDA &4.38 &0.249 \\
CORALl+ \cite{lee19b} &3.95 &0.217 \\
LIP \cite{garcia14a} &3.58 &0.195 \\
\hline
\end{tabular}
\label{tab:exp1}
\vspace*{-4mm}
\end{table}

InD PLDA resulted in lower EER but higher $\rm min$$\rm C_{primary}$ than did OOD PLDA.
We reckon that InD PLDA did not outperform OOD PLDA because the training data for InD PLDA was limited,
so as to be able to train a good PLDA. 
Both domain adaptation methods outperformed any single OOD or InD system. 
It is also expected that the supervised linear interpolation would outperform unsupervised CORAL+.

Performance of the proposed correlation-alignment-based interpolation (CIP) and covariance regularization (reg) is shown in Table~\ref{tab:exp2}.
The weights for all the interpolations were also chosen to be $0.5$. 
CIP performed worse in terms of EER than other methods but achieved a better $\rm min$$\rm C_{primary}$ than the conventional linear interpolation (LIP).
All of the proposed methods performed better than LIP in terms of $\rm min$$\rm C_{primary}$.
The best system is CIP reg domain adaptation which reduced $\rm min$$\rm C_{primary}$ by $41.0\%$ and $30.5\%$,
respectively, as compared with the single InD and OOD system in Table~\ref{tab:exp1}. 
It was also lower by $11.3\%$ than that of LIP. 
\begin{table}[t] 
\centering
\caption{Comparison of LIP and CIP with and without regularization using interpolation weights of $0.5$.}
\begin{tabular}[t]{l|cc}
\hline
Systems & EER(\%) & $\rm min$$\rm C_{primary}$ \\
\hline \hline
LIP \cite{garcia14a} &3.58 &0.195 \\
LIP reg &3.58 &0.195 \\
CIP &3.68 &0.186 \\
CIP reg &3.58 &0.173 \\
\hline
\end{tabular}
\label{tab:exp2}
\vspace*{-3mm}
\end{table}

\begin{figure}[t]
\centering
\includegraphics[width=0.5\textwidth]{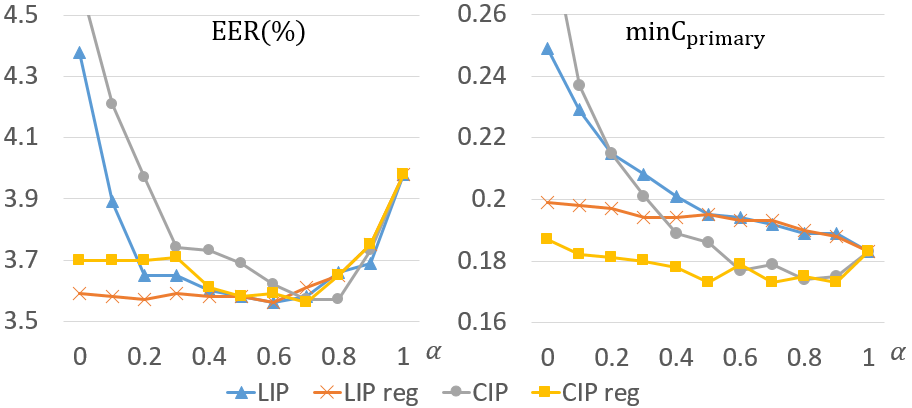}
\vspace*{-5mm}
\caption{The proposed methods with varying weights.}
\label{fig:fig2}
\vspace*{-3mm}
\end{figure}


\begin{figure}[t]
\centering
\includegraphics[width=0.48\textwidth]{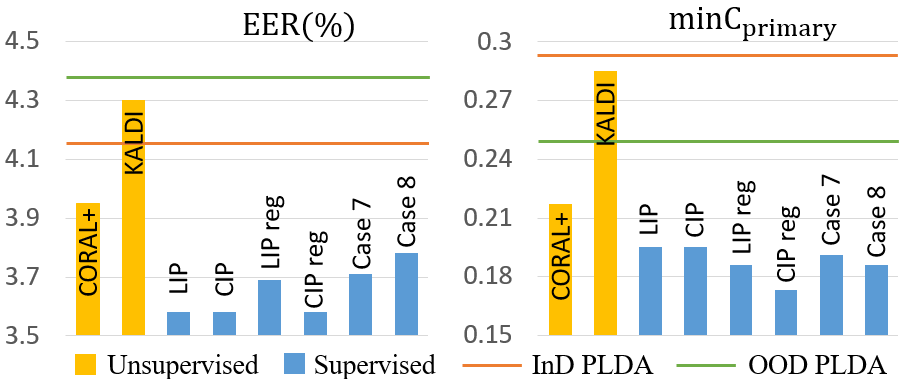}
\vspace*{-5mm}
\caption{Results of special cases derived from the generalized framework using interpolation weights of $0.5$.}
\label{fig:fig3}
\vspace*{-5mm}
\end{figure}

We further investigated the effects on speaker verification performance of varying interpolation weights from $0.0$ to $1.0$ (see Figure~\ref{fig:fig2}).
It can be seen that the proposed covariance regularization technique provided more robust performance 
for both LIP and CIP over a wider range of the interpolation weights. 
This would be beneficial in practice.
For the proposed correlation-alignment-based interpolation (CIP), though its $\rm{EER}$ was worse than that of other interpolations at the weight $\alpha=0.5$ (Table~\ref{tab:exp2}), 
its best $\rm{EER}$ was $3.57\%$ at the weight $\alpha=0.8$, which is comparable to the other three systems' best $3.56\% ~\rm{EER}$. 
Also, the correlation alignment interpolations (CIP and CIP reg) were better than the linear interpolations (LIP and LIP reg) in terms of $\rm min$$\rm C_{primary}$ with all weights.
The best $\rm{EER}$ of the CIP reg system was $5.5\%$ lower than LIP's best. 

Figure~\ref{fig:fig3} summarizes experimental results of all the special cases shown in Table~\ref{tab:generalform1}.
Performance improvement is observed in all cases.

\section{Summary}
\label{sec:summary}

We have proposed here a generalized framework for domain adaptation of PLDA in speaker recognition that works with both unsupervised and supervised methods, 
as well as two new techniques: (1) correlation-alignment-based interpolation and (2) covariance regularization.
The generalized framework enable us to combine the two techniques and also several existing supervised and unsupervised domain adaptation methods into a single formulation. 
Use of the proposed correlation-alignment-based interpolation method decreases $\rm min$$\rm C_{primary}$ up to 30.5\% as compared to that with the out-of-domain PLDA model before adaptation. 
It is also 5.5\% lower than with the conventional linear interpolation method with optimal interpolation weights. 
Further, the proposed regularization technique ensures robustness for interpolations w.r.t. varying interpolation weights, which in practice is essential.


\bibliographystyle{IEEEbib}
\bibliography{strings,refs}

\end{document}